\documentclass[aps,showpacs,amsmath,amssymb,twocolumn,nofootinbib]{revtex4-1}

\usepackage{graphicx}
\usepackage{dcolumn}
\usepackage{bm}
\usepackage{color}
\usepackage[colorlinks=true,pdfstartview=FitV,linkcolor=blue,citecolor=blue,urlcolor=blue]{hyperref}
\usepackage[mathlines]{lineno}
\usepackage[dvipsnames]{xcolor}
\usepackage{amsmath}

\everymath{\displaystyle}
\begin{document}

\newcommand{\up}[1]{$^{#1}$}
\newcommand{\down}[1]{$_{#1}$}
\newcommand{\powero}[1]{\mbox{10$^{#1}$}}
\newcommand{\powert}[2]{\mbox{#2$\times$10$^{#1}$}}

\newcommand{\evm}{\mbox{\rm{eV\,$c^{-2}$}}}
\newcommand{\mevm}{\mbox{\rm{MeV\,$c^{-2}$}}}
\newcommand{\gevm}{\mbox{\rm{GeV\,$c^{-2}$}}}
\newcommand{\pgd}{\mbox{g$^{-1}$\,d$^{-1}$}}
\newcommand{\um}{\mbox{$\mu$m}}
\newcommand{\spix}{\mbox{$\sigma_{\rm pix}$}}
\newcommand{\pav}{\mbox{$\langle p \rangle$}}

\newcommand{\sige}{\mbox{$\bar{\sigma_e}$}}
\newcommand{\mass}{\mbox{$m_\chi$}}
\newcommand{\crystal}{\mbox{$f_\textnormal{c}(q,E_e)$}}
\newcommand{\electron}{\mbox{$\rm{e^-}$}}

\newcommand{\beq}{\begin{equation}}
\newcommand{\eeq}{\end{equation}}
\newcommand{\beqs}{\begin{eqnarray}}
\newcommand{\eeqs}{\end{eqnarray}}

\title{Solving the strong CP problem with horizontal gauge symmetry}

\author{Gongjun Choi,$^{1}$}
\thanks{gongjun.choi@gmail.com}

\author{Tsutomu T. Yanagida,$^{1,2}$}
\thanks{tsutomu.tyanagida@ipmu.jp}

\affiliation{$^{1}$ Tsung-Dao Lee 
Institute, School of Physics and Astronomy, Shanghai Jiao Tong University, Shanghai 200240, China}
\affiliation{$^{2}$ Kavli IPMU (WPI), UTIAS, The University of Tokyo,
5-1-5 Kashiwanoha, Kashiwa, Chiba 277-8583, Japan}
\date{\today}

\begin{abstract}

We present a solution to the strong CP problem, which relies on the horizontal gauge symmetry and CP invariance in a full theory. Similar to other Nelson-Barr type solutions, CP violation in both the strong and weak sectors in the Standard model (SM) is attributed to the condensation of complex scalars $\Phi$ in the model. The model is differentiated by others in that it explains the hierarchy in quark-Higgs Yukawa coupling in the SM based on a series of sequential breaking of the horizontal $SU(3)_{f}$ gauge symmetry. The experimental constraint $\overline{\theta}\lesssim10^{-10}$ requires $<\!\!\Phi\!\!>\,\lesssim\,10^{13}-10^{14}{\rm GeV}$ (vacuum expectation value of complex scalars) and $\lambda\,\lesssim\,10^{-6}$ (scalar quartic coupling). We show that this small coupling is natural in the sense of 'tHooft naturalness. Compared to other models of Nelson-Barr type with CP breaking scale $\Lambda_{CP}\lesssim10^{8}{\rm GeV}$, our model is more advantageous in terms of consistency with the thermal leptogenesis.
\end{abstract}

\maketitle
\section{Introduction}  
A smallness of a parameter in a theory can be considered natural provided certain additional symmetries are restored in the limit where the parameter is sent to zero \cite{tHooft:1979rat}. This sense of naturalness, however, finds an unnatural small parameter when applied to QCD sector of the Standard Model (SM), i.e., $\overline{\theta}=\theta_{0}+{\rm Arg\,det}(Y_{u}Y_{d})$. Here $\theta_{0}$ is the QCD vacuum angle parametrized by a coefficient of the term $\sim F_{\mu\nu}\tilde{F}^{\mu\nu}$ in the QCD sector and $Y_{q}$ is the Yukawa coupling matrix. The parameter enters  in the expression of the neutron electric dipole moment (NEDM) $d_{n}=3.6\times10^{-16}\theta\,{\rm e\,cm}$ \cite{Crewther:1979pi} of which the current experimental constraint $d_{n}<3\times10^{-26}\,{\rm e\,cm}$ \cite{Baker:2006ts}  yields $\overline{\theta}<10^{-10}$. Setting $\overline{\theta}=0$ still does not undo the breaking of CP symmetry because of non-zero KM phase in the weak sector in the SM. 

Several explanations as to the smallness of $\overline{\theta}$ have been suggested. These include, for instance, possibilities of having massless up-quark and the idea of introducing a global $U(1)$ symmetry with a color anomaly \cite{Peccei:1977hh,Peccei:1977ur,Weinberg:1977ma,Wilczek:1977pj}. For the purpose of setting $\overline{\theta}\!=\!0$, the field redefinition of the up-quark and the vacuum expectation value (VEV) of a pseudo Nambu-Goldstone boson arising from the breaking of the anomalous global $U(1)$ can be used for the former and later cases, respectively. The lattice computation of the up-quark mass shows significant deviation from zero \cite{Aoki:2016frl} and thus this simplest solution seems less likely (however, see e.g. \cite{Bardeen:2018fej}). For the later solution \cite{Peccei:1977hh,Peccei:1977ur,Weinberg:1977ma,Wilczek:1977pj,Kim:1979if,Shifman:1979if,Dine:1981rt,Zhitnitsky:1980tq}, a variety of the experimental searches for the pseudo Nambu-Goldstone boson, axion, have  been suggested and performed, and are still under active scrutiny to date (see, e.g. \cite{Graham:2015ouw}).

Another class of solution concerns spontaneous breaking of CP symmetry. The most well known among this line of solutions is the Nelson-Barr model \cite{Nelson:1983zb,Barr:1984qx}. The model begins with the assumption that CP transformation is a symmetry of the model, giving rise to $\theta_{0}=0$. Furthermore, all the interaction coefficients in the Lagrangian become real and the model is constructed in a way that the determinant of the fermion mass matrix is rendered real as long as CP is conserved. The model assumes a complex scalar sector and the vacuum thereof breaks CP. It is VEV of this complex scalar which makes the next leading order contribution to the fermion mass matrix complex, thereby inducing CP violating KM phase in the weak sector of the SM. One of features that makes the solution of this kind distinguished from others is that CP violations in the strong and weak sector are attributed to fundamentally identical physics. 

Along with an unknown fundamental origin of CP violating parameters, i.e. $\overline{\theta}$ and KM phase, an underlying physics responsible for the fermion mass hierarchy remains mysterious in the SM as well. On the other hand, KM phase and the mass hierarchy have something to do with each other in that both are associated with Yukawa coupling matrices in the SM. Given this situation, should one is aimed to resolve the strong CP problem by relying on physics of the spontaneous CP violation, it could be a natural suspicion that underlying origins of $\overline{\theta}$, KM phase and the hierarchy in fermion masses may possibly be very closely related to one another 

In this work, as an answer to such a suspicion, we present a model which contains new heavy quarks and complex scalars apart from the SM particle contents. Also, we extend the gauge group of the SM by introducing horizontal $SU(3)_{f}$ gauge symmetry and by assuming three $Z_{2}$ discrete gauge symmetries. The complex scalars obtain VEVs of different scales, which results in not only the spontaneous breaking of CP symmetry but a series of sequential breaking of $SU(3)_{f}$. Within the model, Yukawa coupling structure of the SM is explained as well as the smallness of $\overline{\theta}$.

\section{Model}

Apart from the SM gauge group, the model has a horizontal (flavor) gauge symmetry $SU(3)_{f}$ as an additional gauge group \cite{Maehara:1978ts,Wilczek:1978xi,Yanagida:1979gs} and three discrete gauge symmetries $Z^{(1)}_{2}\times Z^{(2)}_{2}\times Z^{(3)}_{2}$. The particle contents of the model are extended by introducing into the SM particles three complex scalars $\Phi_{\alpha i}$ and six heavy Dirac fermions, $\Psi^{u}_{\alpha}$ and $\Psi^{d}_{\alpha}$ where $\alpha$ and $i$ are indices for $SU(3)_{f}$ and $Z_{2}^{(i)}$ respectively ($\alpha,i=1,2,3$). The quantum numbers of the particle contents of the model are presented in Table \ref{table:qn}. We see there is no gauge anomaly if we introduce the lepton sector with three right-handed neutrinos \cite{Yanagida:1979gs}.

Within the model, a gauged CP symmetry \cite{Choi:1992xp,Dine:1992ya} is assumed, resulting in real interaction coefficients and $\theta_{0}=0\,\,({\rm or}\,\,\pi)$. On the acquisition of VEVs of the complex scalars $\Phi_{i}$, both CP and $SU(3)_{f}$ become spontaneously broken. \footnote{To avoid the domain wall problem arising from breaking of CP, we assume that CP violation precedes the inflation.} Without loss of generality, we can write down the vacuum of the scalar sector as
\beq
\Phi_{1}=\begin{bmatrix}0\\0\\X_{1}\end{bmatrix}\,,\quad\Phi_{2}=\begin{bmatrix}0\\Y_{2}\\X_{2}\end{bmatrix}\,,\quad\Phi_{3}=\begin{bmatrix}Z_{3}\\Y_{3}\\X_{3}\end{bmatrix}\,,
\label{eq:VEVs}
\eeq

where $X_{1}$ and $Y_{2}$ are real, and the rest is complex. It is assumed that $V_{1}>V_{2}>V_{3}$ where $|\Phi_{i}|\equiv V_{i}$. 

\begin{table}
\centering
\begin{tabular}{|c||c|c|c|c|c|c|c|} \hline
 & $SU(3)_{c}$ & $SU(2)_{L}$ & $U(1)_{Y}$ & $ SU(3)_{f} $ & $Z_{2}^{(1)}$ & $Z_{2}^{(1)}$ &  $Z_{2}^{(1)}$ \\
\hline\hline
$q$      &  $\square$  &  $\square$  & $+1/6$ & $\square$ & + & + & +  \\
$\overline{u}$  &  $\overline{\square}$  &  1  & $-2/3$ & $\overline{\square}$ & + & + & +   \\
$\overline{d}$ &  $\overline{\square}$  &  1  & $+1/3$ & $\overline{\square}$ & + & + & +   \\
$H$      &  1  &  $\square$  & $-1/2$ & 1 & + & + & +   \\
$U$      & $\square$ &  1  & $+2/3$ & $\square$ & + & + & +  \\
$\overline{U}$ & $\overline{\square}$  &  1  & $-2/3$ & $\overline{\square}$ & + & + & +   \\
$D$      &  $\square$  & 1 & $-1/3$ & $\square$ & + & + & +  \\
$\overline{D}$ &  $\overline{\square}$  & 1 & $+1/3$ & $\overline{\square}$ & + & + & +   \\
$\Phi_{1}$ & 1  & 1 & $0$ & $\square$ & - & + & +  \\
$\Phi_{2}$ & 1  & 1 & $0$ & $\square$ & + & - & + \\
$\Phi_{3}$ & 1  & 1 & $0$ & $\square$ & + & + & - \\
\hline
\end{tabular}
\caption{Quantum numbers of the scalar bosons and fermions of the model. $U(D)$ and $\overline{U}(\overline{D})$ are the Weyl spinors to form Dirac fermions $\Psi^{u}=(U,\overline{U}^{*})^{T}$ and $\Psi^{d}=(D,\overline{D}^{*})^{T}$. A subscript of a complex scalar indicates under which $Z_{2}$ the field is odd. Notice that quantum numbers of $\overline{u}$ and $\overline{U}$ are completely identical. We define $\overline{U}$ as the partner of $U$ for the mass term and $\overline{u}$ as an orthogonal direction to $\overline{U}$. The same applies for $\overline{d}$ and $\overline{D}$ too.}
\label{table:qn} 
\end{table}

The renormalizable $SU(3)_{c}\times SU(2)_{L}\times U(1)_{Y}\times SU(3)_{f}\times(Z_{2})^{3}$ invariant Yukawa coupling of the model reads
\beq
\mathcal{L}_{Yuk}=\mathcal{L}_{q}+\mathcal{L}_{Q}+\mathcal{L}_{qQ}\,,
\eeq
with
\beq
\mathcal{L}_{q} = a^{u}H^{\dagger}q_{\alpha}\overline{u}_{\alpha}+a^{d}Hq_{\alpha}\overline{d}_{\alpha}+{\rm h.c.}\,,
\label{eq:LQ}
\eeq
\beq
\mathcal{L}_{Q}=M^{U}U_{\alpha}\overline{U}_{\alpha}+M^{D}D_{\alpha}\overline{D}_{\alpha}+{\rm h.c.}\,,
\label{eq:Lpsi}
\eeq
and
\beq
\mathcal{L}_{qQ}=b^{u}H^{\dagger}q_{\alpha}\overline{U}_{\alpha}+b^{d}Hq_{\alpha}\overline{D}_{\alpha}+{\rm h.c.}\,,
\label{eq:LQpsi}
\eeq
where the interaction coefficients are real due to CP invariance and $SU(2)_{L}$ indices are suppressed. Similar models are considered in \cite{Masiero:1998yi,Evans:2011wj}. Then mass matrices of each of up and down sector fermions in the model become
\beq
\mathcal{M}^{u}=\begin{bmatrix}
\mathcal{M}^{u}_{11} & \mathcal{M}^{u}_{12} \\
\mathcal{M}^{u}_{21} & \mathcal{M}^{u}_{22} 
\end{bmatrix}=\begin{bmatrix}
a^{u}H^{*}_{0}I_{3\times3} & b^{u}H^{*}_{0}I_{3\times3} \\
0 & M^{U}I_{3\times3} 
\end{bmatrix} \,,
\label{eq:massmatrix1}
\eeq
\beq
\mathcal{M}^{d}=\begin{bmatrix}
\mathcal{M}^{d}_{11} & \mathcal{M}^{d}_{12} \\
\mathcal{M}^{d}_{21} & \mathcal{M}^{d}_{22} 
\end{bmatrix}=\begin{bmatrix}
a^{d}H_{0}I_{3\times3} & b^{d}H_{0}I_{3\times3} \\
0 & M^{D}I_{3\times3} 
\end{bmatrix} \,,
\label{eq:massmatrix2}
\eeq
where $\mathcal{M}^{u}$ and $\mathcal{M}^{d}$ are made of four $3\times3$ block matrices and $H_{0}$ is the neutral component of the Higgs $SU(2)_{L}$ doublet. At the electroweak symmetry breaking (EWSB) vacuum, $H_{0}$ has a VEV $|\!\!<\!\!H_{0}\!\!>\!\!|\simeq246{\rm GeV}$. The block matrices shown in Eq.~(\ref{eq:massmatrix1}) and (\ref{eq:massmatrix2}) are proportional to the identity because of $SU(3)_{f}$. At the renormalizable and tree level, the complex scalars $\Phi_{i}$ as a source of CP violation do not appear in the fermion mass matrices since the fermions are even under discrete symmetries while the scalars are odd. 

The determinant of the fermion mass matrix reads \cite{2011arXiv1112.4379P}
\beq
{\rm det}\mathcal{M}=[{\rm det}\mathcal{M}_{11}][{\rm det}(\mathcal{M}_{22}-\mathcal{M}_{21}\mathcal{M}^{-1}_{11}\mathcal{M}_{12})] \,,
\label{eq:Mudet}
\eeq
\beq
\qquad\,\,\,=[{\rm det}\mathcal{M}_{22}][{\rm det}(\mathcal{M}_{11}-\mathcal{M}_{12}\mathcal{M}^{-1}_{22}\mathcal{M}_{21})] \,.
\label{eq:Mudet2}
\eeq
The complex phase of $<\!\!H_{0}\!\!>$ gets cancelled in the product $\mathcal{M}^{u}\mathcal{M}^{d}$. Along with Eq.~(\ref{eq:massmatrix1}) and (\ref{eq:massmatrix2}), this shows that det$\mathcal{M}^{u}\mathcal{M}^{d}$ is real at the tree level. In the next section, we examine additional contributions to the fermion mass matrices which spoil the reality of det$\mathcal{M}^{u}\mathcal{M}^{d}$. 

\section{Non-zero contribution to $\overline{\theta}$}
In this section, we investigate non-zero contribution to $\overline{\theta}$ that arises as a consequence of the structure of the model. RG evolution of $\overline{\theta}$ is negligible since non-vanishing contribution to $\beta$-function of $\overline{\theta}$ takes place at 7-loop order \cite{Ellis:1978hq}. With this, we apply the experimental constraint $\overline{\theta}\lesssim10^{-10}$ from measurement of the NEDM to the energy scale for breaking of CP and $SU(3)_{f}$. This will constrain the VEV of $\Phi_{i}$ and quartic couplings of the complex scalars in the model. 

\subsection{Contribution by higher dimensional operators}
\label{sec:irroperators}
The interaction between the complex scalars and fermions in the model might be induced by a UV physics, which can be studied by Planck-suppressed higher dimensional operators. In this section, we probe possible higher dimensional operators allowed by symmetries in the model. Then we figure out which operators potentially spoil reality of det$\mathcal{M}^{u}\mathcal{M}^{d}$. Those dangerous operators are to be used to constrain vacuum of the scalar sector. 

We start with the dimension 5 operators. For the up-sector, the operators contributing to the mass matrices after the complex scalar condensation are 
\beqs
\mathcal{O}^{(u,5)}_{21}&=&\sum_{i=1}^{3}c^{(u,5)}_{1,i}\frac{\Phi^{\dagger}_{\beta i}\Phi_{\beta i}}{M_{P}}U_{\alpha}\overline{u}_{\alpha}+\sum_{i=1}^{3}c^{(u,5)}_{2,i}\frac{\Phi^{\dagger}_{\alpha i}\Phi_{\beta i}}{M_{P}}U_{\alpha}\overline{u}_{\beta}\cr\cr
&&+\sum_{i=1}^{3}c^{(u,5)}_{3,i}\frac{H^{\dagger}H}{M_{P}}U_{\alpha}\overline{u}_{\alpha} \,,
\label{eq:Ou521}
\eeqs
\beqs
\mathcal{O}^{(u,5)}_{22}&=&\sum_{i=1}^{3}d^{(u,5)}_{1,i}\frac{\Phi^{\dagger}_{\beta i}\Phi_{\beta i}}{M_{P}}U_{\alpha}\overline{U}_{\alpha}+\sum_{i=1}^{3}d^{(u,5)}_{2,i}\frac{\Phi^{\dagger}_{\alpha i}\Phi_{\beta i}}{M_{P}}U_{\alpha}\overline{U}_{\beta}\cr\cr
&&+\sum_{i=1}^{3}d^{(u,5)}_{3,i}\frac{H^{\dagger}H}{M_{P}}U_{\alpha}\overline{U}_{\alpha} \,,
\label{eq:Ou522}
\eeqs
where $M_{P}\simeq2.4\times10^{18}{\rm GeV}$ is the reduced Planck mass and the repeated indices for $SU(3)_{f}$ are assumed to be summed. Here a subscript of an operator indicates the block matrix position in $\mathcal{M}^{u}$ to which the operator contributes (see Eq.~(\ref{eq:massmatrix1})). The superscript specifies up-sector and the operator mass dimension. The coefficients of operators are real due to CP invariance. Contributions to the each mass matrix by dimension 5 operators are hermitian as can be seen in Eq.~(\ref{eq:Ou521}) and Eq.~(\ref{eq:Ou522}). The same operators can be found for the down-sector with $U$ and $\overline{u}$ replaced with $D$ and $\overline{d}$. Note that $\mathcal{M}_{11}$ and $\mathcal{M}_{12}$ are identity matrices up to the dimension 5 operator level. Hermiticity is maintained under addition and inversion, and thus it can be inferred from Eq.~(\ref{eq:Mudet}) that the reality of det$\mathcal{M}^{u}\mathcal{M}^{d}$ remains protected up to dimension 5 operator level. Put another way, Arg(det$\mathcal{M}^{u}\mathcal{M}^{d}$)\,=\,0 (or $\pi$) holds up to dimension 5 operator level.\footnote{$\overline{\theta}\simeq0$ may be favored by observations \cite{Crewther:1979pi}.}

Next, the dimension 6 operators allowed by symmetries of the model are given as

\beqs
\mathcal{O}^{(u,6)}_{11}&=&\sum_{i=1}^{3}a^{(u,6)}_{1,i}\frac{\Phi^{\dagger}_{\beta i}\Phi_{\beta i}}{M_{P}^{2}}H^{\dagger}q_{\alpha}\overline{u}_{\alpha}\cr\cr
&&+\sum_{i=1}^{3}a^{(u,6)}_{2,i}\frac{\Phi^{\dagger}_{\alpha i}\Phi_{\beta i}}{M_{P}^{2}}H^{\dagger}q_{\alpha}\overline{u}_{\beta}\cr\cr
&&+\sum_{i=1}^{3}a^{(u,6)}_{3,i}\frac{H^{\dagger}H}{M_{P}^{2}}H^{\dagger}q_{\alpha}\overline{u}_{\alpha} \,,
\label{eq:Ou611}
\eeqs
\beqs
\mathcal{O}^{(u,6)}_{12}&=&\sum_{i=1}^{3}b^{(u,6)}_{1,i}\frac{\Phi^{\dagger}_{\beta i}\Phi_{\beta i}}{M_{P}^{2}}H^{\dagger}q_{\alpha}\overline{U}_{\alpha}\cr\cr
&&+\sum_{i=1}^{3}b^{(u,6)}_{2,i}\frac{\Phi^{\dagger}_{\alpha i}\Phi_{\beta i}}{M_{P}^{2}}H^{\dagger}q_{\alpha}\overline{U}_{\beta}\cr\cr
&&+\sum_{i=1}^{3}b^{(u,6)}_{3,i}\frac{H^{\dagger}H}{M_{P}^{2}}H^{\dagger}q_{\alpha}\overline{U}_{\alpha} \,,
\label{eq:Ou612}
\eeqs
Again the coefficients of operators are real due to CP invariance. Together with these contributions, $\mathcal{M}^{u}_{11}$ and $\mathcal{M}^{u}_{12}$ are no longer proportional to identity, but become hermitian. In general, a product of hermitian matrices is hermitian only when component hermitian matrices commute each other. The Wilson coefficients of operators in Eq.~(\ref{eq:Ou611}) and Eq.~(\ref{eq:Ou612}) are arbitrary unknowns and thus it remains undecided whether the block matrices in Eq.~(\ref{eq:massmatrix1}), $\mathcal{M}^{u}_{rs}$ ($r,s=1,2$), commute each other. Therefore, we conclude that breaking of the reality of det$\mathcal{M}^{u}\mathcal{M}^{d}$ starts from dimension 6 operator level. 

By using Eq.~(\ref{eq:d5condition}) and Eq.~(\ref{eq:tyukawa}) of which the details would be discussed in the coming discussion in Sec.~\ref{sec:effectiveYukawa}, application of the experimental constraint $\overline{\theta}\lesssim10^{-10}$ to a ratio of leading contributions to an imaginary and a real part of det$\mathcal{M}$ in Eq.~(\ref{eq:Mudet2}) yields
\beqs
\delta\overline{\theta}\,\,\,&&\sim a^{(d,6)}_{2}b^{(d,6)}_{2}c^{(d,5)}_{2}b^{d}\frac{|\Phi_{1}|^{2}|\Phi_{2}|^{2}|\Phi_{3}|^{2}}{M^{D}M_{P}^{5}}\times|y_{b}|^{-3}\cr\cr
&&\simeq a^{(d,6)}_{2}b^{(d,6)}_{2}10^{-20}\times|y_{b}|^{-2}<\!\!<10^{-10}\,,
\label{eq:dim6thetabar}
\eeqs
where we used 
\beq
|y_{b}|\simeq b^{d}c_{2,i=1}^{(d,5)}\frac{|X_{1}|^{2}}{M^{D}M_{P}}\,.
\label{eq:byukawa}
\eeq
The value $10^{-20}$ is estimated from VEVs of the complex scalars which will be obtained in the next discussion about dimension 7 operator contribution to $\overline{\theta}$. The dominant contribution to $\delta\overline{\theta}$ by the down-sector is traced to $|y_{b}|^{-2}>|y_{t}|^{-2}$. From Eq.~(\ref{eq:dim6thetabar}), we can infer that non-zero $\delta\overline{\theta}$ arising from fermion mass matrices up to dimension 6 operators causes CP violation albeit not large enough to be used for constraining parameters in the model. 

Non-zero contributions to Arg(det$\mathcal{M}^{u}\mathcal{M}^{d}$) can also occur in the following part of dimension 7 operators
\beqs
\mathcal{O}^{(u,7)}_{21}&&\ni\sum_{i,j=1}^{3}c^{(u,7)}_{1,ij}\frac{\Phi^{\dagger}_{\gamma i}\Phi_{\gamma i}\Phi^{\dagger}_{\alpha j}\Phi_{\beta j}}{M_{P}^{3}}U_{\alpha}\overline{u}_{\beta}\cr\cr
&&+\sum_{i,j=1}^{3}c^{(u,7)}_{2,ij}\frac{\Phi^{\dagger}_{\gamma i}\Phi_{\gamma j}\Phi^{\dagger}_{\alpha i}\Phi_{\beta j}}{M_{P}^{3}}U_{\alpha}\overline{u}_{\beta}\cr\cr
&&+\sum_{i,j=1}^{3}c^{(u,7)}_{3,ij}\frac{\Phi^{\dagger}_{\gamma i}\Phi_{\gamma j}\Phi^{\dagger}_{\alpha j}\Phi_{\beta i}}{M_{P}^{3}}U_{\alpha}\overline{u}_{\beta}\,,
\label{eq:Ou721}
\eeqs
\beqs
\mathcal{O}^{(u,7)}_{22}&\ni&\sum_{i,j=1}^{3}d^{(u,7)}_{1,ij}\frac{\Phi^{\dagger}_{\gamma i}\Phi_{\gamma i}\Phi^{\dagger}_{\alpha j}\Phi_{\beta j}}{M_{P}^{3}}U_{\alpha}\overline{U}_{\beta}\cr\cr
&&+\sum_{i,j=1}^{3}d^{(u,7)}_{2,ij}\frac{\Phi^{\dagger}_{\gamma i}\Phi_{\gamma j}\Phi^{\dagger}_{\alpha i}\Phi_{\beta j}}{M_{P}^{3}}U_{\alpha}\overline{U}_{\beta}\cr\cr
&&+\sum_{i,j=1}^{3}d^{(u,7)}_{3,ij}\frac{\Phi^{\dagger}_{\gamma i}\Phi_{\gamma j}\Phi^{\dagger}_{\alpha j}\Phi_{\beta i}}{M_{P}^{3}}U_{\alpha}\overline{U}_{\beta}\,,
\label{eq:Ou722}
\eeqs
where the repeated indices for $SU(3)_{f}$ are assumed to be summed. Corresponding operators of the similar form can be found in the down-sector. In Eq.~(\ref{eq:Ou721}) and Eq.~(\ref{eq:Ou722}), we only showed $\boldsymbol{1}\otimes\boldsymbol{8}\otimes\boldsymbol{8}$ type operators although there are three more types of operators including $\boldsymbol{8}\otimes\boldsymbol{8}\otimes\boldsymbol{8}$, $\boldsymbol{8}\otimes\boldsymbol{8}\otimes\boldsymbol{1}$ and $\boldsymbol{1}\otimes\boldsymbol{1}\otimes\boldsymbol{1}$. For the current purpose of estimating an order of magnitude for $\delta\overline{\theta}$ due to dimension 7 operators, it suffices to study $\boldsymbol{1}\otimes\boldsymbol{8}\otimes\boldsymbol{8}$ type operators below. Up to dimension 7 operator level, the up-sector fermion mass matrix is given by 
\beqs
\mathcal{M}^{u}&&=\begin{bmatrix}
\mathcal{M}^{u}_{11} & \mathcal{M}^{u}_{12} \\
\mathcal{M}^{u}_{21} & \mathcal{M}^{u}_{22} 
\end{bmatrix}\cr\cr&&=\begin{bmatrix}
a^{u}H^{*}_{0}I_{3\times3}\!+\!\mathcal{O}_{11}^{(u,6)}\!&\!\!b^{u}H^{*}_{0}I_{3\times3}\!+\!\mathcal{O}_{12}^{(u,6)} \\
\!\!\!\!\!\!\!\!\mathcal{O}_{21}^{(u,5)}\!\!+\!\mathcal{O}_{21}^{(u,7)}\!\!\!\!\!\!&\!\!\!\!\!\!M^{U}I_{3\times3}\!+\!\mathcal{O}_{22}^{(u,5)}\!+\!\mathcal{O}_{22}^{(u,7)} 
\end{bmatrix} \,.
\label{eq:Mudim7}
\eeqs
For dimension 7 operator level, differing from the previous lower dimensional cases, it is realized that there occur irremovable complex phases in the diagonal elements of $\mathcal{M}^{u}_{21}$ and $\mathcal{M}^{u}_{22}$. Namely, $\mathcal{M}^{u}_{21}$ and $\mathcal{M}^{u}_{22}$ are no longer hermitian matrices. This is because $c^{(u,7)}_{2,ij}= c^{(u,7)}_{2,ji}$, $c^{(u,7)}_{3,ij}= c^{(u,7)}_{3,ji}$, $d^{(u,7)}_{2,ij}= d^{(u,7)}_{2,ji}$ and $d^{(u,7)}_{3,ij}= d^{(u,7)}_{3,ji}$ are not ensured for the second and the third type contributions to $\mathcal{O}^{(u,7)}_{21}$ and $\mathcal{O}^{(u,7)}_{22}$ in Eq.~(\ref{eq:Ou721}) and (\ref{eq:Ou722}). The same applies for the down-sector. Now this fact makes complexity of det$\mathcal{M}^{u}\mathcal{M}^{d}$ manifest.

For estimation of $\delta\overline{\theta}$ arising at the level of the dimension 7 operators, we compare the dominant contributions to a real part and an imaginary part of det$\mathcal{M}^{u}\mathcal{M}^{d}$ by referring to Eq.~(\ref{eq:Mudet2}). Up to dimension 5 operator level, the dominant contribution to the real part comes from det$\mathcal{M}_{21}$, which reads $\sim (|y_{t}|M^{U})^{3}$. On the other hand, we found the dominant contribution to the imaginary part to be $\sim b^{u}(|y_{t}|M^{U})^{2}\mathcal{O}_{21}^{(u,7)}$. The same applies for the down-sector. Then, the ratio of these two produces $\delta\overline{\theta}\sim (b^{u}\mathcal{O}_{21}^{(u,7)})/(|y_{t}|M^{U})+(b^{d}\mathcal{O}_{21}^{(d,7)})/(|y_{b}|M^{D})$. Thus, we obtain
\beq
\delta\overline{\theta}\,\,\sim\, \frac{b^{q}c^{(q,7)}_{2,i=1,j=2}}{|y_{q_{3}}|}\frac{|X_{1}|^{2}|X_{2}|^{2}}{M^{Q}M_{P}^{3}}\lesssim10^{-10}\,,
\label{eq:VEVconstraint}
\eeq
where Eq.~(\ref{eq:VEVconstraint}) shows a greater contribution among up and down-sector, and $q_{3}$ is either of $t$ or $b$. With the use of Eq.~(\ref{eq:tyukawa}), we can rewrite Eq.~(\ref{eq:VEVconstraint}) as 
\beq
\delta\overline{\theta}\,\,\sim\,\frac{c^{(q,7)}_{2,i=1,j=2}}{c^{(q,5)}_{2}}\frac{|X_{2}|^{2}}{M_{P}^{2}}\lesssim10^{-10}\,,
\eeq
where $q$ can be either of $u$ or $d$, depending on which is making a greater contribution. With $c^{(q,7)}_{2,i=1,j=2}/c^{(q,5)}_{2}=10^{P}$ taken, we obtain the upper bound on $|X_{2}|\lesssim10^{(26-P)/2}{\rm GeV}$. For instance, for $P=0$ and $P=-2$, the upper bound reads $10^{13}{\rm GeV}$ and $10^{14}{\rm GeV}$, respectively.

Now that we obtain the upper bound of $|X_{2}|$ in terms of values of the Wilson coefficients of the dimension 5 and 7 operators, we realize that $\delta\overline{\theta}$ due to dimension 6 operators hardly exceeds $10^{-10}$ unless we have fine-tuned Wilson coefficients for dimension 7 operators. Hence, we conclude that $\overline{\theta}\lesssim10^{-10}$ constrains VEVs of the complex scalars at dimension 7 operator level. Yet, CP violation starts at dimension 6 operator level.

Before ending this section, it is worth reconsidering the physical reason for breaking of reality of det$\mathcal{M}^{u}\mathcal{M}^{d}$ at dimension 6 operator level. In other words, why does CP get violated especially from dimension 6 operator level? The fact that $\mathcal{M}^{q}_{11}$ and $\mathcal{M}^{q}_{12}$ $(q=u,d)$ remain proportional to identity was the reason to make det$\mathcal{M}^{u}\mathcal{M}^{d}$ real up to dimension 5 operator level (see Eq.~(\ref{eq:Mudet})). This was possible due to the horizontal $SU(3)_{f}$ gauge symmetry. Also, the fact that the heavy fermions including $\overline{U}$ are $SU(2)_{L}$ singlet disallows dimension 5 operator contribution to $\mathcal{M}_{12}$. In this way, we may understand that the reality of det$\mathcal{M}^{u}\mathcal{M}^{d}$ up to dimension 5 operator level is related to the horizontal $SU(3)_{f}$ gauge symmetry and $SU(2)_{L}$ singlet heavy fermions. Nonetheless, $\mathcal{M}^{q}_{11}$ and $\mathcal{M}^{q}_{12}$ $(q=u,d)$ are no longer proportional to the identity matrix starting from the dimension 6 operator level, resulting in non-zero contribution to $\overline{\theta}$.

\subsection{One loop contribution to $\bar{\theta}$}
\label{sec:oneloop} 
In the previous section, we observed tree-level non-vanishing contribution to $\bar{\theta}$ arises from dimension 7 operators, which constrains the symmetry breaking scale for both CP and the $SU(3)_{f}$ horizontal gauge symmetry. In this section, we will investigate how the scalar sector within the model is constrained by one loop radiative correction to $\overline{\theta}$. To this end, we begin with the following renormalizable scalar potential which respects $SU(3)_{f}\times (Z_{2})^{3}$,
\beqs
V(\Phi)&=&-\sum_{i=1}^{3}\frac{1}{2}m_{\Phi_{i}}^{2}|\Phi_{\alpha i}|^{2}+\frac{\lambda_{0}}{4}\sum_{i,j=1}^{3}|\Phi_{\alpha i}|^{2}|\Phi_{\beta j}|^{2}\cr\cr
&&+\frac{\lambda_{+}}{4}\sum_{i,j=1}^{3}\Phi^{\dagger}_{\alpha i}\Phi_{\alpha j}\Phi^{\dagger}_{\beta i}\Phi_{\beta j}\cr\cr
&&+\frac{\lambda_{-}}{4}\sum_{i,j=1}^{3}\Phi^{\dagger}_{\alpha i}\Phi_{\alpha j}\Phi^{\dagger}_{\beta j}\Phi_{\beta i}\cr\cr
&&+\,\,{\rm higher\,\, order\,\, terms\,,}
\label{eq:Vphi}
\eeqs
where the repeated $SU(3)_{f}$ indices are assumed to be summed. Notice that the higher order terms are non-negligible to determine all VEVs for $\Phi_{i}$, since $\lambda_{\pm}$ are very small as shown below. The determination of VEVs of $\Phi_{i}$ is beyond the scope of this paper. Again the CP invariance renders all the interaction coefficients in $V(\Phi)$ in Eq.~(\ref{eq:Vphi}) real.

For contributions to $\delta\overline{\theta}$ by higher dimensional operators, we observed a significant imaginary part of det$\mathcal{M}^{u}\mathcal{M}^{d}$ arises at the dimension 7 operator level. And also we observed the dominant contribution to the imaginary part of det$\mathcal{M}^{u}\mathcal{M}^{d}$ is attributable to det$\mathcal{M}_{21}^{q}$ ($q=u,d$) in Eq.~(\ref{eq:Mudet2}). This implies that the leading radiative correction to det$\mathcal{M}_{21}^{q}$ ($q=u,d$) with four scalar condensation external lines must be also constrained by the experimental constraint $\overline{\theta}\lesssim10^{-10}$.

With that being said, our aim is to ascertain whether one loop corrections to the block matrices of $\mathcal{M}^{u}_{21}$ and $\mathcal{M}^{d}_{21}$ spoil the hermiticity after the condensation of $\Phi$ and thus induce Arg(det$\mathcal{M}^{u}\mathcal{M}^{d}$)$\neq0$. If the hermiticity is broken in $\mathcal{M}_{21}^{q}$, then the loop correction must be constrained. Recall that $c^{(q,7)}_{2,ij}\neq c^{(q,7)}_{2,ji}$ and $c^{(q,7)}_{3,ij}\neq c^{(q,7)}_{3,ji}$ ($q=u,d$) were the essential points to make diagonal components of $\mathcal{M}^{u}_{21}$ and $\mathcal{M}^{d}_{21}$ complex when we discussed dimension 7 operators (see the second and third type contributions to $\mathcal{O}^{(u,7)}_{21}$ in Eq.~(\ref{eq:Ou721})). Hence, testing hermiticity at the loop level reduces to checking whether the loop corrections to $c^{(q,7)}_{2,ij}$ and $c^{(q,7)}_{2,ji}$ ($c^{(q,7)}_{3,ij}$ and $c^{(q,7)}_{3,ji}$) are identical or not. 

For each loop correction to $c^{(q,7)}_{2,ij}$ ($c^{(q,7)}_{3,ij}$), if there exists a corresponding identical correction to $c^{(q,7)}_{2,ji}$ ($c^{(q,7)}_{3,ji}$) up to a loop factor, then the one loop corrections do not break hermiticity of $\mathcal{M}^{u}_{21}$ and $\mathcal{M}^{d}_{21}$ by accomplishing $\delta c^{(q,7)}_{2,ij}= \delta c^{(q,7)}_{2,ji}$ ($\delta c^{(q,7)}_{3,ij}= \delta c^{(q,7)}_{3,ji}$). As a matter of fact, we find that this is not the case by observing breaking of the one to one correspondence. In the following, we demonstrate this by showing a correction to $c^{(q,7)}_{2,ij}$ does not have its partner correction to $c^{(q,7)}_{2,ji}$. The same thing can be observed for $c^{(q,7)}_{3,ij}$ and $c^{(q,7)}_{3,ji}$ as well.

\begin{figure}[h]
\centering
\hspace*{-5mm}
\includegraphics[width=0.55\textwidth]{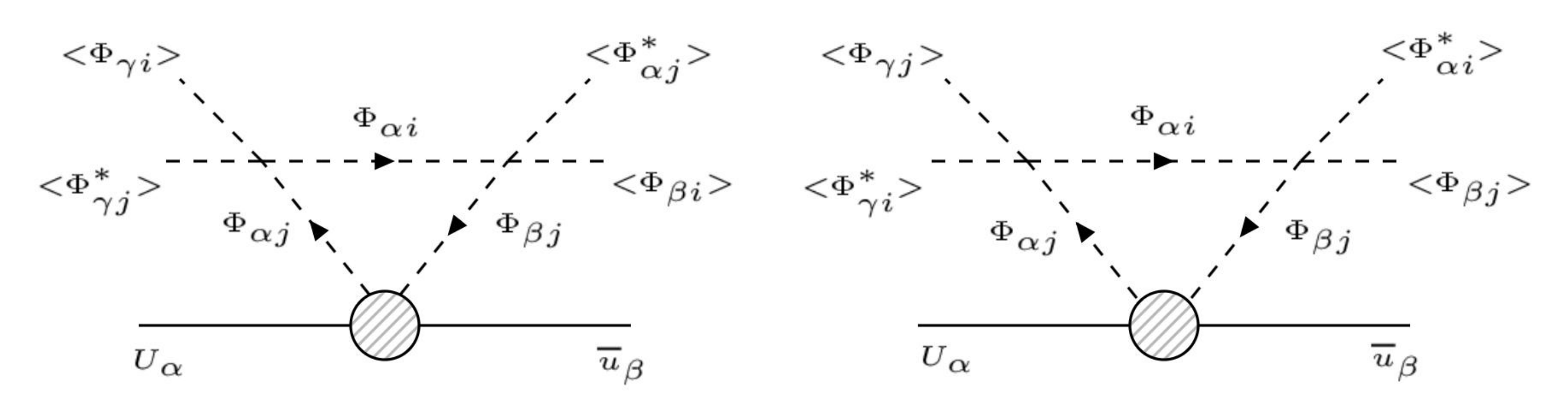}
\caption{One loop correction to the dimension 7 operator with coefficients $c^{(u,7)}_{2,ij}$. The right panel diagram corresponds to a correction of the dimension 7 operator with coefficients $c^{(u,7)}_{2,ij}$. The left panel diagram corresponds to a correction of the dimension 7 operator with coefficients $c^{(u,7)}_{2,ji}$. The blobs correspond to the dimension 5 operators with the coefficient $c^{(u,5)}_{2}$. The arrow on the internal line is directed from $\Phi^{\dagger}$ to $\Phi$.}
\vspace*{-1.5mm}
\label{fig:1}
\end{figure}

Among many different contributions, in Fig.~\ref{fig:1} we show loop corrections to $c^{(u,7)}_{2,ij}$ and $c^{(u,7)}_{2,ji}$ with the same internal lines in the right and left panel, respectively. Provided the scalar quartic vertex factors are different, then we may conclude that the loop correction to $c^{(u,7)}_{2,ij}$ in the right panel lacks its identical partner for $c^{(u,7)}_{2,ji}$. The same applies for the down-sector.

In accordance with Eq.~(\ref{eq:Vphi}), we see that vertex factors for scalar quartic interactions read ($\lambda_{-}\lambda_{+}$) and  ($\lambda_{+}\lambda_{0}$) for the left and right panel diagrams respectively in Fig.~\ref{fig:1}. This proves that the one loop corrections to $c^{(q,7)}_{2,ij}$ and $c^{(q,7)}_{2,ji}$ are different and thus there occur complex values on the diagonal components of $\mathcal{M}^{u}_{21}$ and $\mathcal{M}^{d}_{21}$. One may wonder what happens if we modify the kind of scalar fields in the internal lines such that we can have the same scalar quartic interaction factors for the left and right panel diagrams in Fig.~\ref{fig:1}. However, in those cases, now the momentum space integral for the loop becomes different because of different scalar masses for different kinds of scalars in internal lines. To prevent the one loop correction from making $\delta\overline{\theta}$ exceed $10^{-10}$, we demand 
\beqs
(\delta\overline{\theta})_{\rm 1-loop}&&\simeq\frac{\delta\mathcal{M}^{q}_{21,\alpha=3\beta=3}}{\mathcal{M}^{q}_{21,\alpha=3\beta=3}}\cr\cr&&\simeq \frac{b^{q}c^{(q,5)}}{|y_{q_{3}}|}\frac{\lambda^{2}}{16\pi^{2}}\frac{|X_{1}|^{2}|X_{2}|^{2}}{m_{\Phi}^{2}M^{Q}M_{P}}\lesssim10^{-10} \,,
\label{eq:looptheta1}
\eeqs
where Eq.~(\ref{eq:looptheta1}) presents a dominant contribution among up and down sector. Now we do not specify the kind of $\lambda$ in Eq.~(\ref{eq:looptheta1}) for simplicity. Using $|X_{1}|=10^{R}|X_{2}|$ and Eq.~(\ref{eq:tyukawa}), we can further simplify Eq.~(\ref{eq:looptheta1}) to obtain
\beq
(\delta\overline{\theta})_{\rm 1-loop}\simeq10^{-2R}\frac{\lambda^{2}}{16\pi^{2}}\frac{|X_{1}|^{2}}{m^{2}_{\Phi}}\lesssim10^{-10}\,,
\label{eq:looptheta2}
\eeq
where $m_{\Phi}$ is the mass of heaviest scalar particle in the loop. For $\lambda^{2}$, there are three possibilities: (1) $\lambda_{0}^{2}$ (2) $\lambda_{0}\lambda_{\pm}$ (3) $\lambda_{\pm}\lambda_{\pm}$. We found that the first case does not violate the hermiticity of one loop corrections to the block matrix $\mathcal{M}^{u}_{21}$. For the rest of two cases, we find $m_{\Phi}^{2}\simeq |X_{1}|^{2}\lambda_{0}$ provided at least one of three internal lines corresponds to a massive scalar mode.\footnote{For $\lambda_{0}>\!\!>\lambda_{\pm}$, the complex scalar $\Phi_{i}$ contains two classes of bosons. One has heavier masses of $|X_{i}|^{2}\lambda_{0}\,(i=1,2,3)$ and the other lighter masses of $|X_{i}|^{2}\lambda_{\pm}\,(i=2,3)$. However, we checked that our conclusion does not change. } Therefore, we may argue that both non-zero $\lambda_{+}$ and $\lambda_{-}$ are responsible for spoiling the hermiticity of one loop corrections to $\overline{\theta}$ and thus subject to the upper bound on $\lambda$ obtained above while constraining $\lambda_{0}$ is not necessary to fulfill $\overline{\theta}\lesssim10^{-10}$. From Eq.~(\ref{eq:looptheta2}), the upper bound on $\lambda_{\pm}$ is obtained to be $\lambda_{\pm}\lesssim10^{-8+2R}$. For an exemplary case with $R=1$, the constraint becomes $\lambda_{\pm}\lesssim10^{-6}$.

\begin{figure}[h]
\centering
\hspace*{-5mm}
\includegraphics[width=0.25\textwidth]{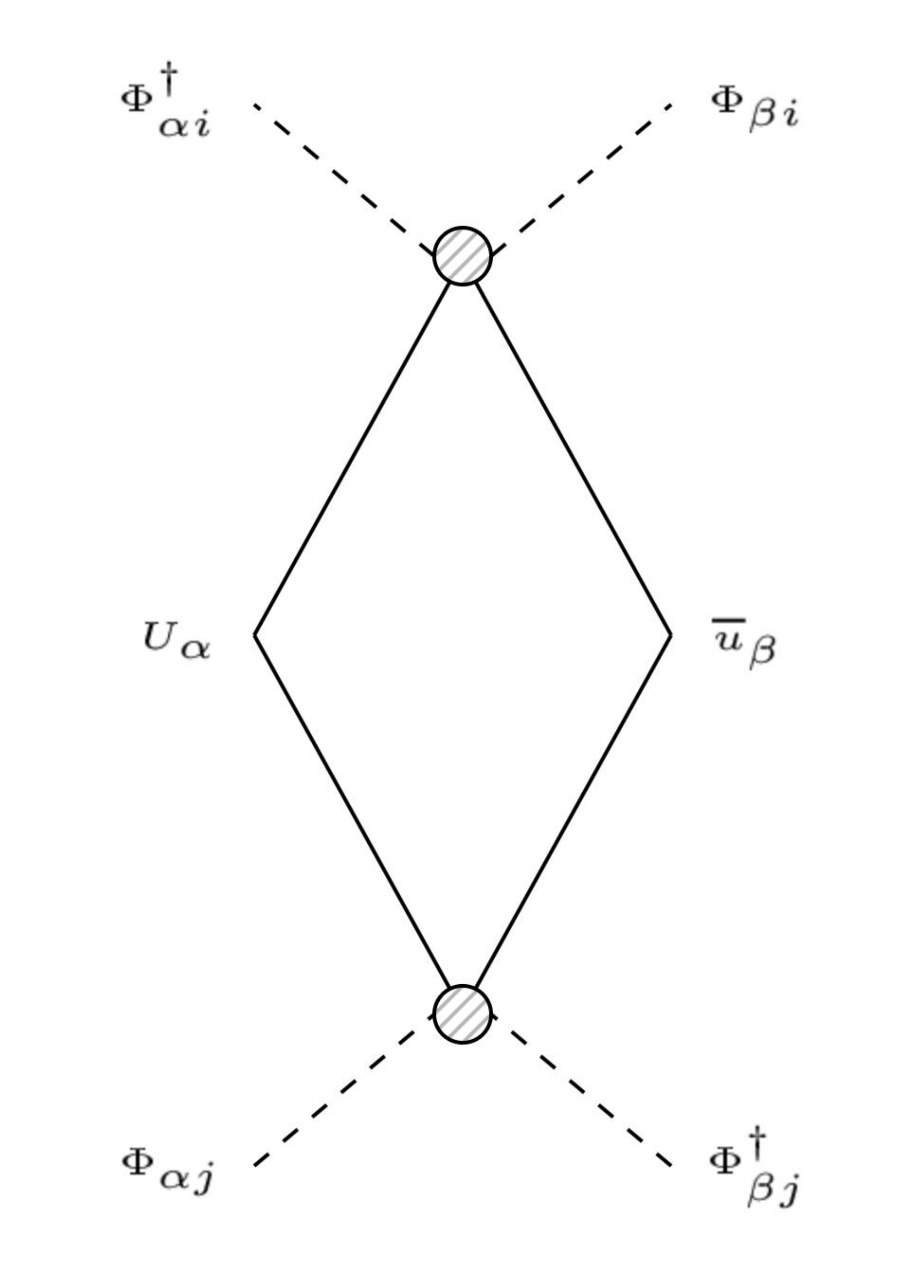}
\caption{A diagram for one loop radiative correction to $\lambda_{-}$ quartic interaction of the complex scalar fields defined in Eq.~(\ref{eq:Vphi}). The blobs correspond to the dimension 5 operator with the coefficient $c_{2,i}^{(u,5)}$ defined in Eq.~(\ref{eq:Ou521}).}
\vspace*{-1.5mm}
\label{fig:2}
\end{figure}

Having $\lambda_{\pm}\lesssim10^{-8+2R}$, we realize the Wilson coefficients of dimension 5 operators we discussed in Sec.~\ref{sec:irroperators} can be constrained. The dimension 5 operators with the coefficients $c_{2}^{(q,5)}$ and $d_{2}^{(q,5)}$ can induce radiative correction to $\lambda_{-}$. For example, the diagram shown in Fig.~{\ref{fig:2}} makes the radiative correction to $\lambda_{-}$ by the amount of $\sim (c_{2,i}^{(u,5)})^{2}/(16\pi^{2})$. Thus, the constraint on $\lambda_{-}$ transforms into $c_{2,i}^{(u,5)} \lesssim10^{(-6+2R)/2}$ and $d_{2}^{(q,5)}\lesssim10^{(-6+2R)/2}$. For instance, $R=1$ case results in $c_{2,i}^{(u,5)}\lesssim10^{-2}$ and $d_{2}^{(q,5)}\lesssim10^{-2}$.

Thus far we have discussed one loop correction to dimension 7 operator of $\boldsymbol{1}\otimes\boldsymbol{8}\otimes\boldsymbol{8}$ type only. However, for $\boldsymbol{1}\otimes\boldsymbol{1}\otimes\boldsymbol{1}$, $\boldsymbol{8}\otimes\boldsymbol{8}\otimes\boldsymbol{1}$ and $\boldsymbol{8}\otimes\boldsymbol{8}\otimes\boldsymbol{8}$ type dimension 7 operators, we found that one loop corrections with four external scalar lines are hermitian whenever those are proportional to $\lambda_{0}^{2}$. Therefore, we conclude that there is no constraint on $\lambda_{0}$ as far as one loop corrections to dimension 7 operators are concerned.

The upper bounds on $\lambda_{\pm}$, $c_{2,i}^{(q,5)}$ and $d_{2,i}^{(q,5)}$ ($q=u,d$) now change the point of view in which we understand the smallness of $\overline{\theta}$. The smallness of $\overline{\theta}$ turns out to be originated from the smallness of $\lambda_{\pm}$, $c_{2,i}^{(q,5)}$ and $d_{2,i}^{(q,5)}$ ($q=u,d$). So at first glance, it seems that what the model achieves is just to convert the form of the smallness. The later smallness, however, can be considered distinguished from the former in that the symmetry of the model gets enhanced to include $SU(3)_{f}^{(1)}\times SU(3)_{f}^{(2)}\times SU(3)_{f}^{(3)}$ in the limit $\lambda_{\pm}$, $c_{2,i}^{(q,5)}$, $d_{2,i}^{(q,5)}\rightarrow0$. Here the superscript on $SU(3)_{f}^{(i)}$ specifies to which the complex scalar field $\Phi_{i}$ $SU(3)_{f}^{(i)}$ applies. This enables us to understand the smallness of $\lambda_{\pm}$, $c_{2,i}^{(q,5)}$ and $d_{2,i}^{(q,5)}$ ($q=u,d$) natural in the sense of 'tHooft \cite{tHooft:1979rat}. In sum, the model succeeds in converting the unnatural smallness of $\overline{\theta}$ into other natural smallness.

\section{Effective Yukawa Coupling}
\label{sec:effectiveYukawa}
In this section, we study how the model can produce the effective Yukawa coupling in the SM. For this purpose, it turns out that we need constraints on Wilson coefficients of higher dimensional operators as we shall see below.

\begin{figure}[t]
\centering
\hspace*{-5mm}
\includegraphics[width=0.5\textwidth]{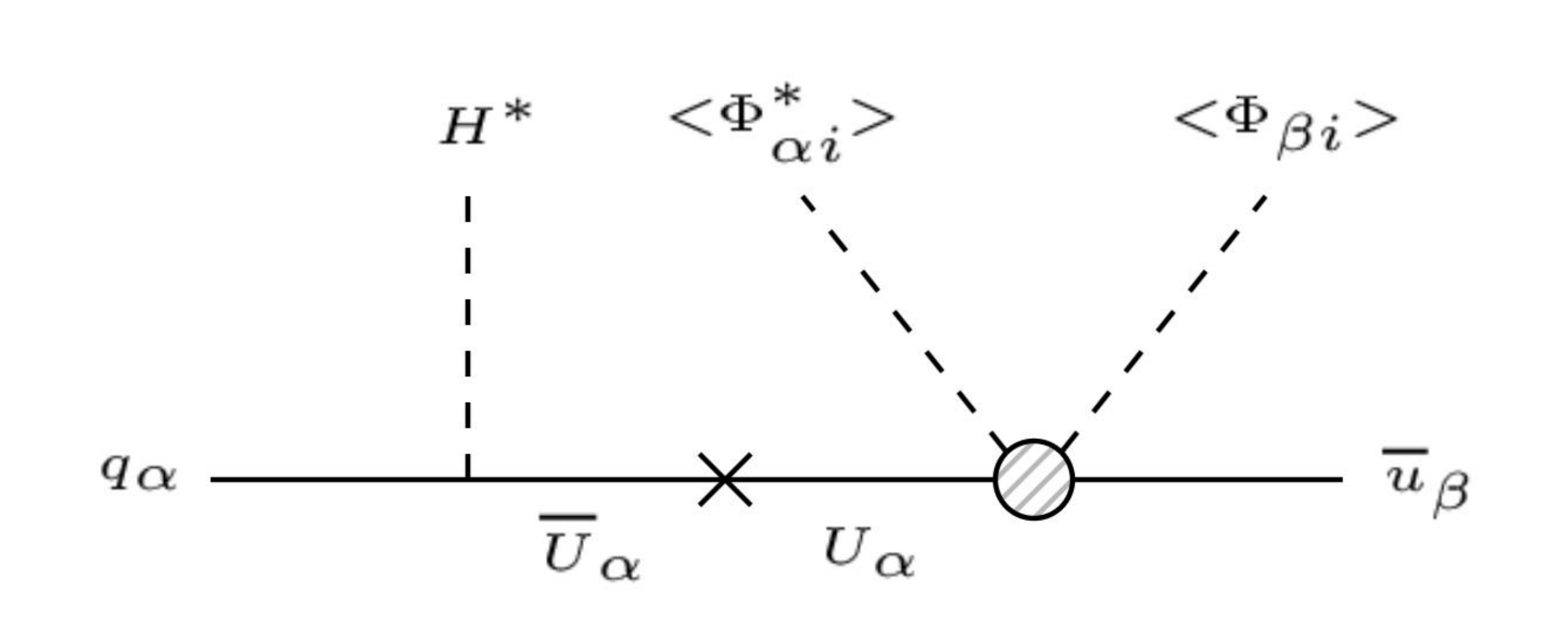}
\caption{Diagram which produces a low energy effective SM quark-Higgs Yukawa coupling after spontaneous breaking of $SU(3)_{f}$ and CP. The sum over $Z_{2}$ index $i$ is assumed. The blob corresponds to the higher dimensional operator in Eq.~(\ref{eq:Ou521}). The cross represents the mass insertion of the heavy fermion.}
\vspace*{-1.5mm}
\label{fig:3}
\end{figure}

If we assume $d^{(q,5)}$ is small enough to make the following condition satisfied
\beq
d^{(q,5)}\frac{\Phi^{\dagger}_{\alpha i}\Phi_{\alpha i}}{M_{P}} <\!\!< M^{Q}\,,
\label{eq:d5condition}
\eeq
where $q=u,d$ and $Q=U,D$, then for the energy scale between the EWSB scale and the heavy fermion mass scale $M^{Q}$, the effective SM quark-Higgs Yukawa coupling can be obtained by integrating out the heavy fermions $\Psi^{u}$ and $\Psi^{d}$. The relevant diagram of a UV physics contribution to the SM Yukawa coupling is shown in Fig.~\ref{fig:3}. With this, the effective Yukawa coupling induced in the low energy reads
\beqs
Y^{u}_{\alpha\beta}&\simeq& a^{u}+\sum_{i=1}^{3} b^{u}c^{(u,5)}_{1,i}\frac{<\!\!\Phi^{\dagger}_{\alpha i}\!\!><\!\!\Phi_{\beta i}>}{M^{U}M_{P}}\delta_{\alpha\beta}\cr\cr&&+\sum_{i=1}^{3} b^{u}c^{(u,5)}_{2,i}\frac{<\!\!\Phi^{\dagger}_{\alpha i}\!\!><\!\!\Phi_{\beta i}>}{M^{U}M_{P}}\,,
\label{eq:effYu}
\eeqs
\beqs
Y^{d}_{\alpha\beta}&\simeq& a^{d}+\sum_{i=1}^{3} b^{d}c^{(d,5)}_{1,i}\frac{<\!\!\Phi^{\dagger}_{\alpha i}\!\!><\!\!\Phi_{\beta i}>}{M^{D}M_{P}}\delta_{\alpha\beta}\cr\cr&&+\sum_{i=1}^{3} b^{d}c^{(d,5)}_{2,i}\frac{<\!\!\Phi^{\dagger}_{\alpha i}\!\!><\!\!\Phi_{\beta i}>}{M^{D}M_{P}}\,,
\label{eq:effYd}
\eeqs
where $a^{q},b^{q}$ $(q=u,d)$  and the heavy fermion masses are defined in Eq.~(\ref{eq:massmatrix1}) and Eq.~(\ref{eq:massmatrix2}), and $c^{(q,5)}$ $(q=u,d)$ is from Eq.~(\ref{eq:Ou521}). The hermitian Yukawa coupling in the low energy turns out to be one of the features of the model.

Depending on a value of $a^{q}$ ($q=u,d$), there can be two different situations for the effective SM Yukawas. In the first place, $a^{q}$ can be $\mathcal{O}(1)$ so as to be the leading contribution. In this case, with other parameters, $a^{q}$ needs to be tuned for reproducing the SM quark masses as a free parameter of the model  \cite{Masiero:1998yi}. Since the structure of the SM Yukawa will be dominated by $a^{q}$ and other contributions, VEVs of $\Phi_{i}$s cannot explain that of the SM Yukawas without the parameter tuning. In the second place, as a free parameter, $a^{q}$ can be small enough to be negligible in comparison with other contributions \cite{Evans:2011wj}. For example, an accidental symmetry can be introduced which is not respected by terms in Eq.~(\ref{eq:LQ}) in order to suppress $a^{q}$. Then, $Y^{q}$s in Eq.~(\ref{eq:effYu}) and Eq.~(\ref{eq:effYd}) become dominated by dimension 5 operators and thus the hierarchical structure can be explained by the hierarchy of VEVs of the complex scalars $\Phi_{i}$. Since $Y^{q}$ given in Eq.~(\ref{eq:effYu}) and Eq.~(\ref{eq:effYd}) are hermitian, $a^{q}$s do not affect CKM matrix for both cases. In our work, we considered the second scenario.

For $\alpha=3,\beta=3$, $Y^{q}$ is matched to the SM top-Higgs and bottom-Higgs Yukawa coupling. Then, for energy scale between EWSB scale and $M^{Q}$, we have
\beq
|y_{q_{3}}|\simeq b^{q}(c_{1,i=1}^{(q,5)}+c_{2,i=1}^{(q,5)})\frac{|X_{1}|^{2}}{M^{Q}M_{P}}\,,
\label{eq:tyukawa}
\eeq
where $q_{3}=t,b$ and $Q=U,D$. Especially for $q_{3}=t$, we demand $|y_{t}|\simeq\mathcal{O}(1)$. Recall that $|X_{1}|=10^{R}|X_{2}|$ and $c_{2,i=1}^{(u,5)}\lesssim10^{(-6+2R)/2}$. If $|b^{u}|\lesssim |y_{t}|$ holds at the scale of $M_{Q}$\footnote{In this case, the mixing between $\overline{t}$ and $\overline{U}_{3}$ becomes non-negligible and thus the observed $\overline{t}^{'}$ is a mixture of $\overline{t}$ and $\overline{U}_{3}$. We need more precise calculation for the up-type quark mass matrix than that estimated below.}, then with the constraint on  $|X_{2}|\lesssim 10^{(26-P)/2}{\rm GeV}$, the assumption that $c_{1,i=1}^{(u,5)}$ and $c_{2,i=1}^{(u,5)}$ are comparable gives $M^{U}\lesssim10^{5-P+3R}{\rm GeV}$.  As an example, for $P=0$ and $R=1$, $M^{U}$ cannot be greater than $10^{8}$ GeV. On the other hand, if $|b^{u}|>|y_{t}|$ holds at the scale of $M_{Q}$, then we see that a larger value of $b^{u}$ gives rise to a larger value of $M^{U}$ to meet Eq.~(\ref{eq:tyukawa}). However, notice that there must be an upper bound for $b^{u}$ at the scale of $M_{Q}$ to avoid breaking of perturbativity at $SU(3)_{f}$ breaking scale. Then, with Eq.~(\ref{eq:d5condition}) and (\ref{eq:tyukawa}), we realize that the condition $d^{(u,5)}<c^{(u,5)}$ should be satisfied even in this case.

Now if a UV physics allows $c_{1,i}^{(q,5)}\lesssim c_{2,i}^{(q,5)}$ $(q=u,d)$, then we may understand the hierarchical structure in the SM Yukawas by hierarchy between the complex scalar VEVs, i.e. $<\!\!\Phi_{1}\!\!>\,>\,<\!\!\Phi_{2}\!\!>\,>\,<\!\!\Phi_{3}\!\!>$. Assuming that conditions $a^{u},a^{d}<\!\!<1$ and $c_{1,i}^{(q,5)}\lesssim c_{2,i}^{(q,5)}$ $(q=u,d)$ can hold in a UV physics, the SM quark mass matrices can be approximately described by the 14 free parameters which appear in the above Eq.~(\ref{eq:effYu}) and Eq.~(\ref{eq:effYd}). Firstly, the ratios of VEVs $\Phi_{2}$ and $\Phi_{3}$ in Eq.~(\ref{eq:VEVs}) to $<\!\!\Phi_{1}\!\!>=X_{1}$ multiplied by the ratio of coefficients produce 9 free real parameters. Namely,
\beq
p_{\alpha j}\equiv\sqrt{\frac{c_{2,j}^{(u,5)}}{c_{2,i=1}^{(u,5)}}}\frac{<\!\!\Phi_{\alpha j}\!\!>}{|X_{1}|}\,,
\eeq
where $(\alpha,j)=(2,2)$ forms one real and $(3,2)$, $(1,3),\,(2,3),\,(3,3)$ do another 4 complex parameters. Secondly, the ratios between Wilson coefficients $c_{2,i}^{(d,5)}/c_{2,i}^{(u,5)}$ with $i=1,2,3$ give additional 3 free parameters. Lastly, the ratios $b^{u}/M^{U}$ and $b^{d}/M^{D}$ give the last two more parameters. The last two are irrelevant for CKM quark mixing matrix, but play a role of scaling factors for each individual quark mass eigenvalues. 

With a curiosity as to the hierarchy in VEVs of complex scalars $\Phi_{i}$, we carried out the procedure to reproduce the 4 quark mass ratios, 3 mixing angles and Jarlskog invariant in the SM. Interestingly, assuming $c_{2,j}^{(u,5)}/c_{2,i=1}^{(u,5)}\sim\mathcal{O}(1)$, the ratios $|Y_{2}|/|X_{1}|\simeq0.08$ and $|X_{2}|/|X_{1}|\simeq0.6$ are obtained.

\section{Conclusions}
In this paper, we present a model as a resolution to the strong CP problem. The SM particle content is extended by including additional complex scalars $\Phi$ and heavy fermions $\Psi^{u}$ ($\Psi^{d}$). Furthermore, the model introduces the horizontal $SU(3)_{f}$ gauge symmetry and discrete gauge symmetries $Z^{(1)}_{2}\times Z^{(2)}_{2}\times Z^{(3)}_{2}$ as additives to the SM gauge group. The quantum numbers of the particle content of the model can be referred to from Table.~\ref{table:qn}.

Beginning as a gauged CP invariant theory, the spontaneous CP violation becomes triggered by the complex scalar field $\Phi_{i}$ condensation. Simultaneously, the horizontal $SU(3)_{f}$ gets spontaneously broken around the energy scale $\sim10^{13}-10^{14}{\rm GeV}$. This is in contrast with other Nelson-Barr type models where the CP breaking occurs for $\Lambda_{CP}\lesssim10^{8}{\rm GeV}$ by dimension 5 operators \cite{Dine:2015jga}. The higher breaking scale of our model is better in avoiding a tension to the thermal leptogenesis \cite{Fukugita:1986hr,Buchmuller:2005eh}. We found that provided the scalar sector of the model is featured by small enough quartic self-interaction at this scale, i.e., $\lambda\lesssim10^{-6}$, then the radiatively induced CP violating parameter in QCD sector, $\overline{\theta}$, can be small enough to avoid the current experimental constraint $\overline{\theta}\lesssim10^{-10}$. The upper bound $\lambda\lesssim10^{-6}$ further constrains the Wilson coefficient of dimension 5 operators to be smaller than $10^{-2}$. The newly obtained smallness of other parameters in the model than $\overline{\theta}$ turns out to be technically natural, enhancing the symmetry of the model \cite{tHooft:1979rat}. 

On the other hand, the quark-Higgs Yukawa coupling structure is explained as a consequence of the model. The sequential breaking (so-called tumbling) of the horizontal $SU(3)_{f}$ gauge symmetry by different VEVs of three complex scalars leads on to the hierarchical structure of the effective Yukawa coupling in the SM \cite{Wilczek:1978xi}. Within the model, CKM matrix is determined by the scalar field sector dynamics and the interplay between fermions and scalars communicated by a UV physics. Therefore, in this work, we find that CP violation in the strong and weak sector, and the hierarchical structure of the Yukawa coupling in the SM are originated from the common underlying physics of breaking of CP and $SU(3)_{f}$ induced by the scalar field dynamics. Extension including the lepton sector will be given elsewhere.


\begin{acknowledgments}

We thanks to Yue Zhao for his collaboration in the early stage of this work and for valuable comments on the draft. T. T. Y. thanks to Kazuya Yonekura for discussion on the QCD vacuum and is supported in part by the China Grant for Talent Scientiﬁc Start-Up Project and the JSPS Grant-in-Aid for Scientiﬁc Research No. 16H02176, No. 17H02878, and No. 19H05810 and by World Premier International Research Center Initiative (WPI Initiative), MEXT, Japan. T. T. Y. thanks to Hamamatsu\!\!\!\!\!\!\!\! Photonics.

\end{acknowledgments}


\bibliography{ref}

\end{document}